\documentclass[PRL,reprint,superscriptaddress,showpacs]{revtex4-1}

\usepackage{amsmath}    
\usepackage{graphicx}   
\usepackage{color}      
\usepackage{subfigure}  
\usepackage{hyperref}   
\usepackage{todonotes}
\usepackage{hyperref}
\usepackage{amssymb}
\usepackage[T1]{fontenc}
\usepackage[utf8]{inputenc}

\begin{document}

\title{Low energy exciton pocket at finite momentum in tetracene molecular solids}
\author{Friedrich Roth}
\affiliation{Center for Free-Electron Laser Science / DESY, Notkestra\ss e 85, D-22607 Hamburg, Germany}
\author{Markus Nohr}
\affiliation{IFW Dresden, P.O. Box 270116, D-01171 Dresden, Germany}
\author{Silke Hampel}
\affiliation{IFW Dresden, P.O. Box 270116, D-01171 Dresden, Germany}
\author{Martin Knupfer}
\affiliation{IFW Dresden, P.O. Box 270116, D-01171 Dresden, Germany}
\date{\today}

\begin{abstract}
The excited state dynamics in organic semiconductors plays an important role for many processes associated with light absorption and emission. We have studied the momentum dependence of the lowest singlet excitons in tetracene molecular solids, an archetype system for other organic semiconductors. Our results reveal an anisotropic bandstructure of these excitons with an energy minimum at finite momentum, i.\,e., a low energy exciton pocket. The existence of such low energy states might have important consequences for the photophysical behavior, also in view of applications in, e.\,g., organic solar cells. Our studies stress the importance of momentum dependent considerations in organic systems.
\end{abstract}

\maketitle

\section{Introduction}
The electronic excitations in molecular crystals have been subject of many investigations through decades \cite{PopeBuch}. This has on the one hand been motivated by fundamental question and interest, on the other hand those crystals formed by $\pi$ conjugated molecules (also called organic semiconductors) additionally promise particular applications in organic (opto-)electronic devices \cite{Walzer2007,Virkar2010,Klauk2010,Scharber2013,Reineke2013,Yu2014}. It is well established that in the majority of organic semiconductors the lowest electronic excitations form excitons (bound electron-hole pairs) \cite{PopeBuch}, whereas the character and dynamics of these excitons are decisive for light absorption and emission with a particular importance for emerging technologies such as organic photovoltaics \cite{Zhugayevych2015}. In this context, investigations of archetype materials such as tetracene offer fundamental insight into exciton dynamics and can thus advance our understanding of many related organic semiconducting materials.

\par

Despite the large number of studies carried out in the past, some features related to the exciton dynamics in tetracene are still unclear \cite{Tayebjee2013,Burdett2011,Burdett2010,Wilson2013}. From temperature dependent and time resolved photoluminescence studies the existence of a dark or ``dull'' state has been proposed, the nature of which is open yet. Also the presence of trap states has been discussed, e.\,g., in relation to temperature dependent energy variations in the luminescence signal. While the vast majority of previous studies addressed the time, temperature and polarization dependent behavior of tetracene and other organic semiconductors, little is known on the momentum dependence. In crystals, excitons are characterized by a momentum $q$ which is linked to their propagation \cite{PopeBuch,DavydovBuch,Cudazzo2015}. This gives rise to an exciton band structure or dispersion E($q$), analogous to other quasi-particles like phonons or electrons. For pentacene, a close relative of tetracene, detailed studies of the exciton dispersion or exciton band structure have provided us with important insight into the nature of the excitons, in particular they have provided clear evidence that the fundamental singlet excitons cannot be rationalized on the basis of molecular Frenkel excitons only. Instead, a significant mixture of Frenkel and charge transfer (CT) excitations has to be considered \cite{Schuster2007,Roth2012}. Moreover, this has also been concluded from theoretical considerations, which show that neither the Davydov splitting, arising from the interaction of the two inequivalent molecules in the unit cell of acenes, nor the exciton dispersion can be understood without the inclusion of charge transfer excitons \cite{Tiago2003,Hummer2005,Yamagata2011,Cudazzo2012,Cudazzo2013,Sharifzadeh2013,Pac2014}.

\par

In this work we present a comprehensive analysis of the singlet exciton dispersion in tetracene along fundamental reciprocal lattice vectors at low temperature using electron energy-loss spectroscopy (EELS) in transmission.EELS is based on the inelastic scattering of fast electrons, whereby both energy and momentum are transferred to the electronic excitations of the materials under investigation (see Ref. \cite{Roth2014} for details). This method is able to determine the energy of electronic excitations at finite momentum transfers throughout the entire Brillouin zone of a material, which can provide substantially more insight into the character of the excitations \cite{Schuster2007,Knupfer1999,Knupfer2002,Kramberger2008}. We demonstrate that the energetically lowest singlet excitons in tetracene are characterized by a sizable and anisotropic dispersion. Moreover, our studies reveal a so far unknown low energy pocket at finite momentum.

\section{Experimental}

We have grown high quality tetracene single crystals via physical transport in an inert gas stream consisting of argon and hydrogen. The starting material, tetracene powder (Sigma-Aldrich Chemie GmbH), was sublimed in the hot zone of a furnace at about 300$^{\circ}$C. Crystals with dimensions of about 10\,mm $\times$ 7\,mm $\times$ 0.1\,mm were grown at a temperature of around 170$^{\circ}$C. The growth lasted 2 to 6 hours. The high crystal quality has been confirmed by x-ray diffraction, infrared- and Raman spectroscopy.

\par

Investigations using EELS in transmission require tetracene films with a thickness of about 100\,nm. These were cut from large, flat surfaces of the single crystals with an ultramicrotome using a diamond knife. Subsequently, the films were mounted onto standard electron microscopy grids and transferred into the EELS spectrometer.  All EELS measurements were carried out using a 170\,keV spectrometer thoroughly discussed in previous publications \cite{Fink1989,Roth2014}. We note that at such high primary beam energy only singlet excitations are possible. The energy and momentum resolution were 85 meV and 0.03\,\AA$^{-1}$, respectively (momentum values are given in reciprocal space directions in units of \AA$^{-1}$). The measured signal represents the electronic excitation spectrum and is proportional to $\operatorname{Im}(-1/\epsilon$) ($\epsilon$ is the energy and momentum dependent dielectric function), and it was determined for various momentum transfers, $q$, parallel to the directions of the corresponding reciprocal lattice vectors. A He flow cryostat allowed to keep the films at a temperature of 20\,K during the measurements.

\par

\begin{figure}[t]
\includegraphics[width=.9\linewidth]{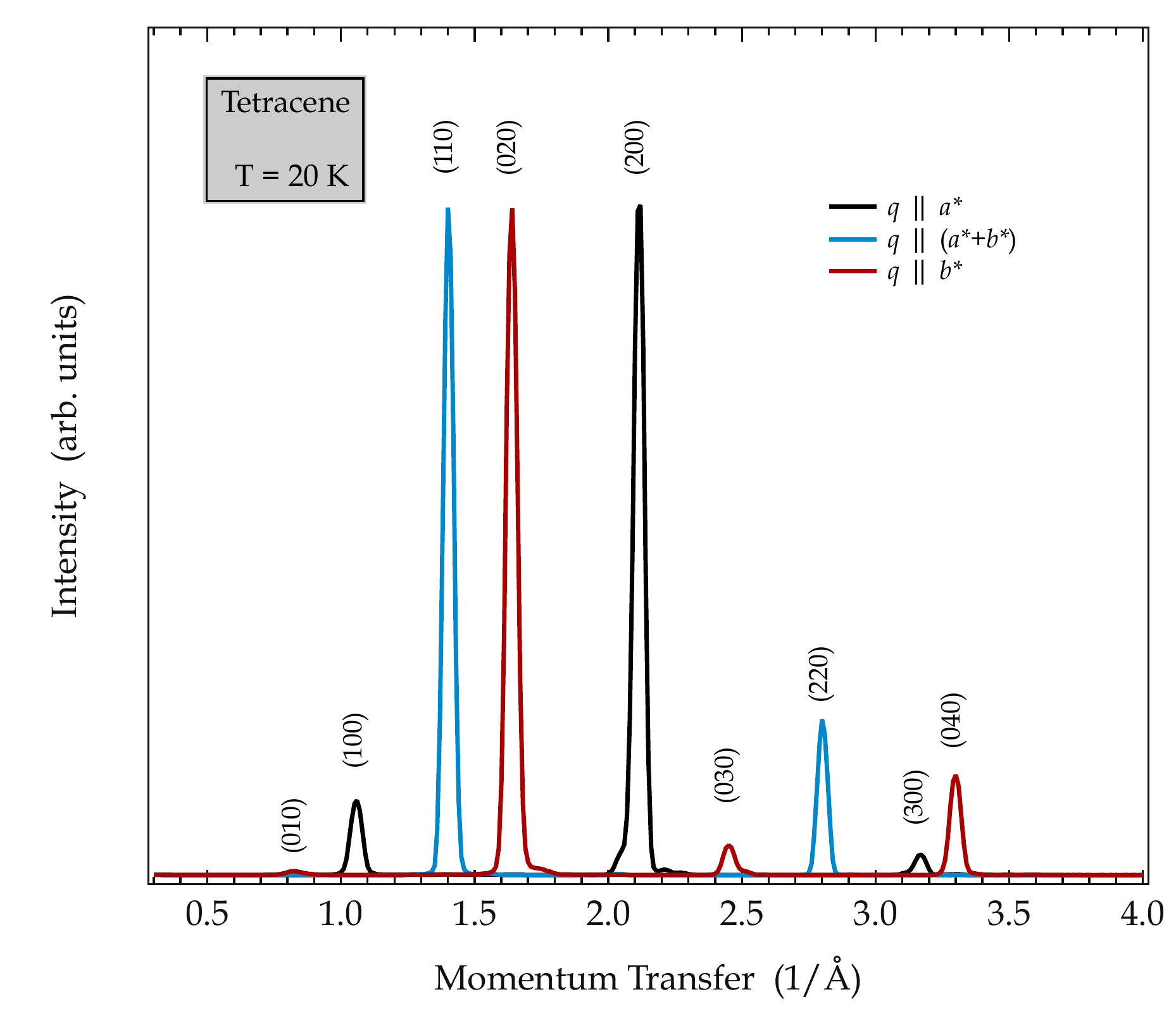}
\caption{Selected electron diffraction profiles at 20\,K of the tetracene films as studied in this contribution. The data represent measurements along fundamental reciprocal lattice directions and reveal the single crystalline nature of the films.}
\label{bragg}
\end{figure}

In order to determine the electronic excitations as a function of momentum transfer in different directions, we have thoroughly characterized our films using \textit{in-situ} electron diffraction (cf. Fig.\,\ref{bragg}). The momentum directions were then selected parallel to the corresponding reciprocal lattice directions. Moreover, the diffraction profiles document that the films are single crystalline as can be seen from Fig.\,\ref{bragg}, where we show the electron diffraction profiles parallel to selected reciprocal lattice directions. These data correspond well to those from X-ray diffraction of our tetracene single crystals as well as to published X-ray diffraction data \cite{Holmes1999}. In this contribution we use the crystal axis notation of Ref. \cite{Cudazzo2015}, where $a^*$ is the longer reciprocal lattice vector.

\par

Since molecular crystals often can be damaged by fast electrons, we checked our samples for any sign of degradation. Sample degradation was followed by watching an increasing amorphous-like background in the electron diffraction spectra and an increase of spectral weight in the excitation spectra in the energy region below the first excitation feature. It turned out that under our measurement conditions the spectra remained unchanged for more than 12 hours. Samples that showed any signature of degradation were not considered further but replaced by freshly prepared thin films. Data from different films were fully reproducible.

\section{Results and discussion}

\begin{figure}[b]
\includegraphics[width=.9\linewidth]{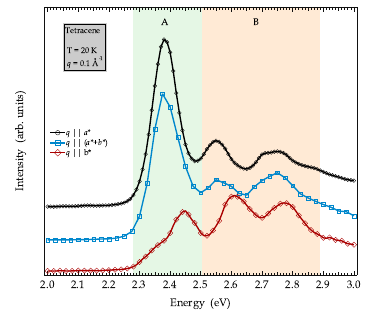}
\caption{Electronic singlet excitations of single crystalline tetracene for a small momentum transfer ($q$ = 0.1\,\AA$^{-1}$) and for momentum vectors $q$ parallel to selected reciprocal lattice directions.The green shaded area A and orange shaded area B display the region where we observe the Davydov components as well as the vibronic satellites, respectively.}
\label{aniso}
\end{figure}

In the molecular crystal, tetracene adopts a triclinic crystal structure with two inequivalent molecules in the unit cell \cite{Holmes1999}. This causes the lowest singlet exciton to split into two Davydov components which renders the electronic excitation spectra significantly anisotropic \cite{PopeBuch,Bree1960}. This particular anisotropy is also present in our data. In Fig.\,\ref{aniso} we compare the excitation spectra for momentum vectors $q$ parallel to selected reciprocal lattice directions and a small absolute value of the momentum transfer of 0.1\,\AA$^{-1}$. At such a small momentum transfer, we probe essentially vertical transitions, the so-called optical limit, i.\,e., a comparison to optical data is possible. Moreover, the direction of our momentum vector $q$ corresponds to that of the light polarization vector in an optical experiment. Our data in Fig.\,\ref{aniso} correspond very well to those from optical data, the observed Davydov splitting is about 70\,meV, and the two components are essentially polarized along the two crystal directions $a$ and $b$ \cite{Bree1960}. However, it is important to realize that our data are taken along the reciprocal lattice directions. Therefore, the spectrum in Fig.\,\ref{aniso} for $q$ parallel to the reciprocal direction $b^*$ has two features around 2.4\,eV, which arise from the two Davydov components, while for a momentum transfer parallel to $a^*$ and ($a^*$+$b^*$) we only observe the lower Davydov feature (cf. Fig.\,\ref{aniso} area A). Furthermore, the curves in Fig.\,\ref{aniso} also reveal higher energy satellites starting above 2.5\,eV (see Fig.\,\ref{aniso} area B). Also the spectral shape and the relative energy of these satellite features are in very good agreement to the results of optical measurements. These satellites have been ascribed to mainly vibronic progressions, but we note that there is also evidence for further electronic excitations in this energy range \cite{Bree1960,Yamagata2011,Tiago2003,Cudazzo2012,Roth2013}. Thus, to disentangle vibronic and electronic contribution and to develop a microscopic understanding in the energy region of the satellites still represents a challenge, and we will concentrate on the exciton band structure of the leading feature in the remainder of this contribution.

\par

\begin{figure}[t]
\includegraphics[width=0.9\linewidth]{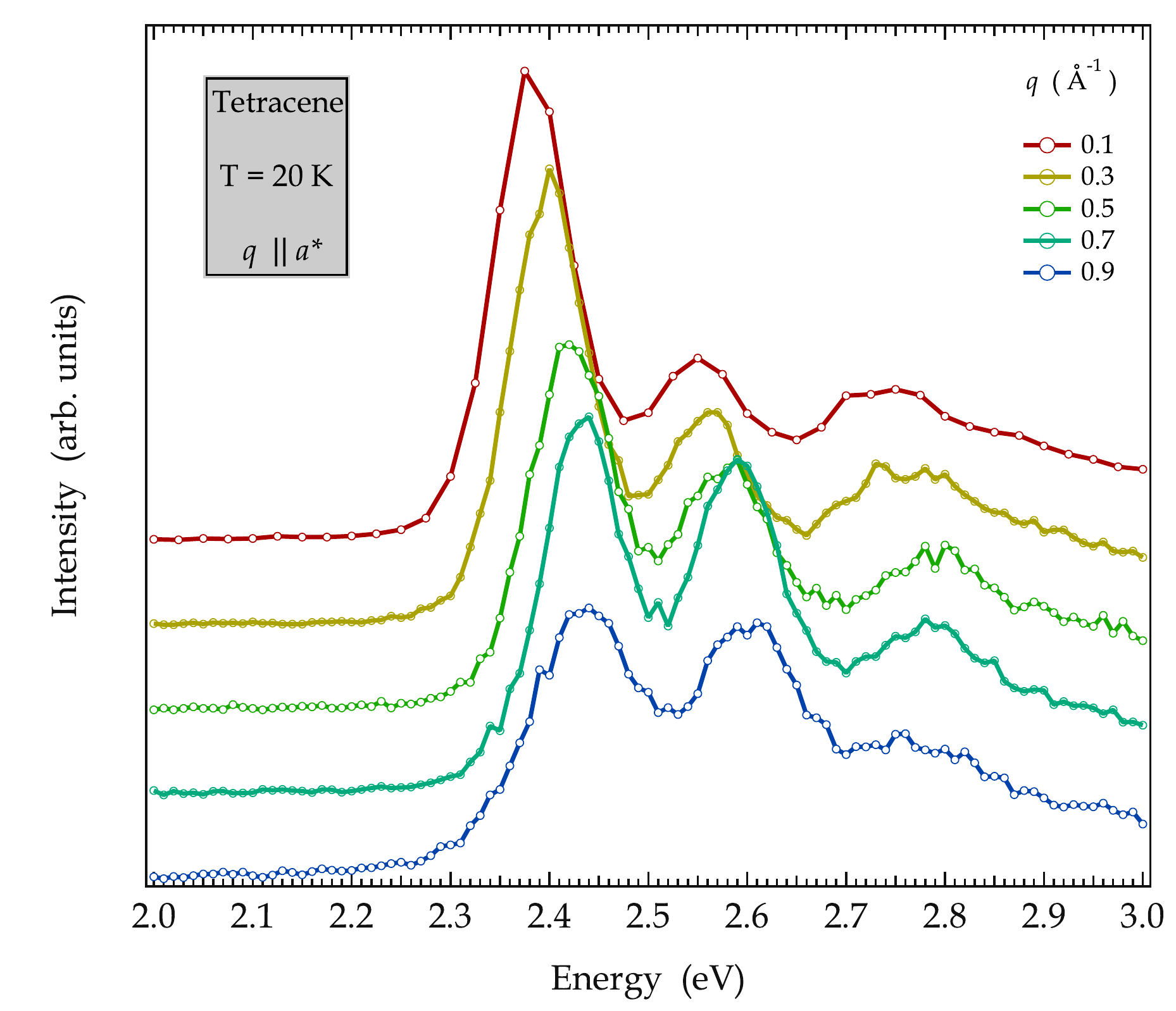}
\includegraphics[width=0.9\linewidth]{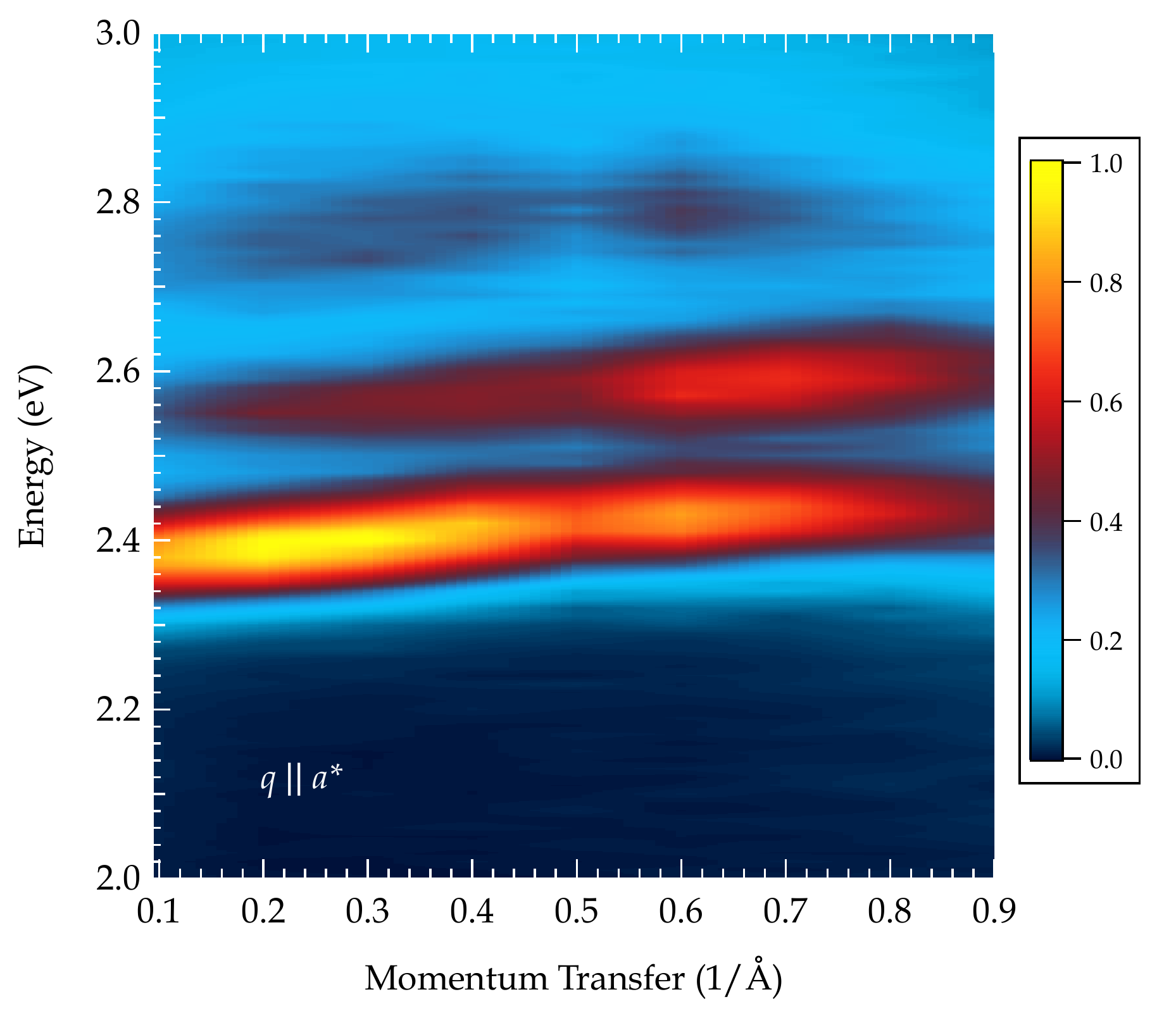}
\caption{Excitation spectra of tetracene as a function of momentum transfer parallel to the reciprocal lattice direction $a^*$.}
\label{dispa}
\end{figure}

\begin{figure}[t]
\includegraphics[width=0.9\linewidth]{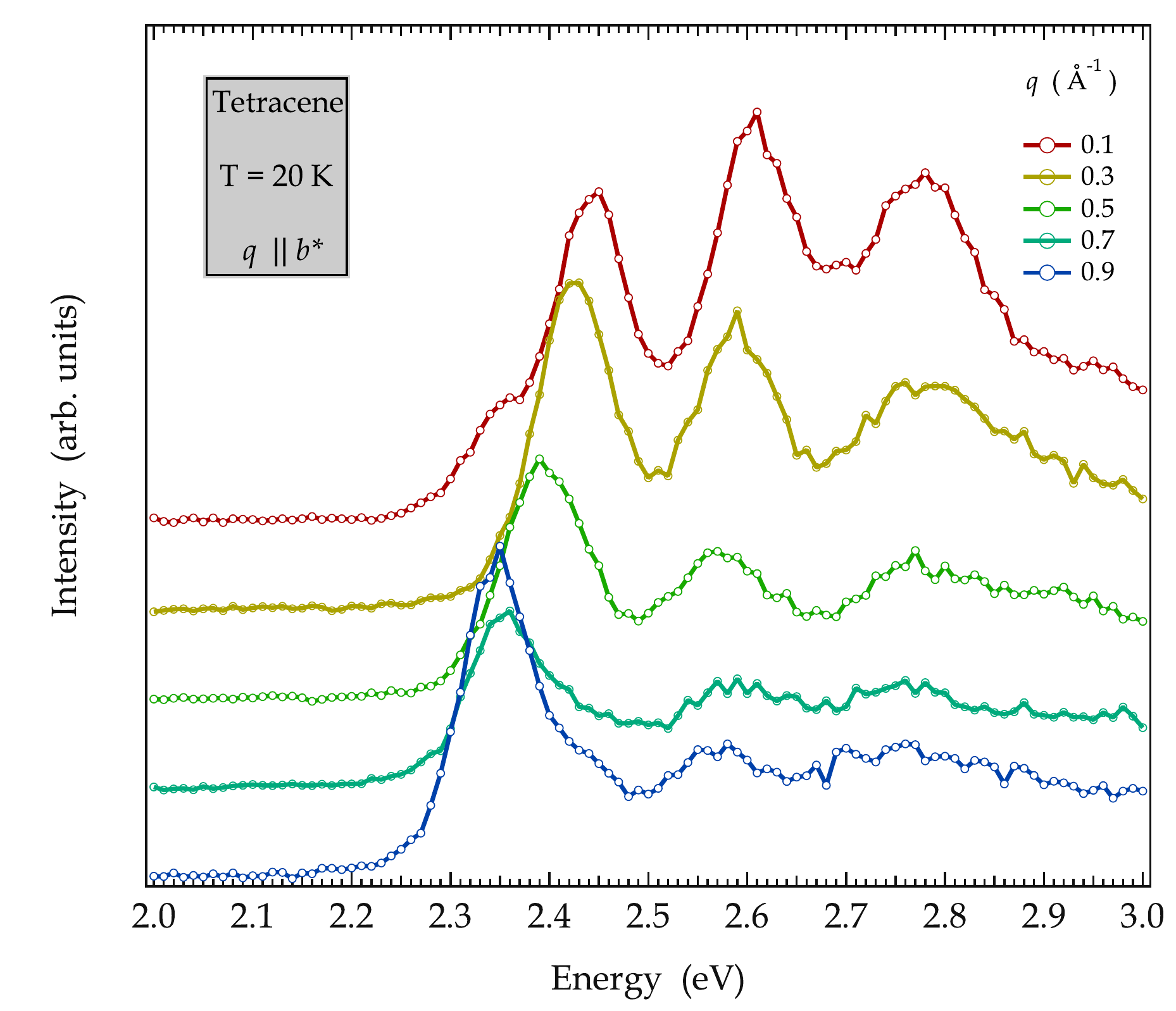}
\includegraphics[width=0.9\linewidth]{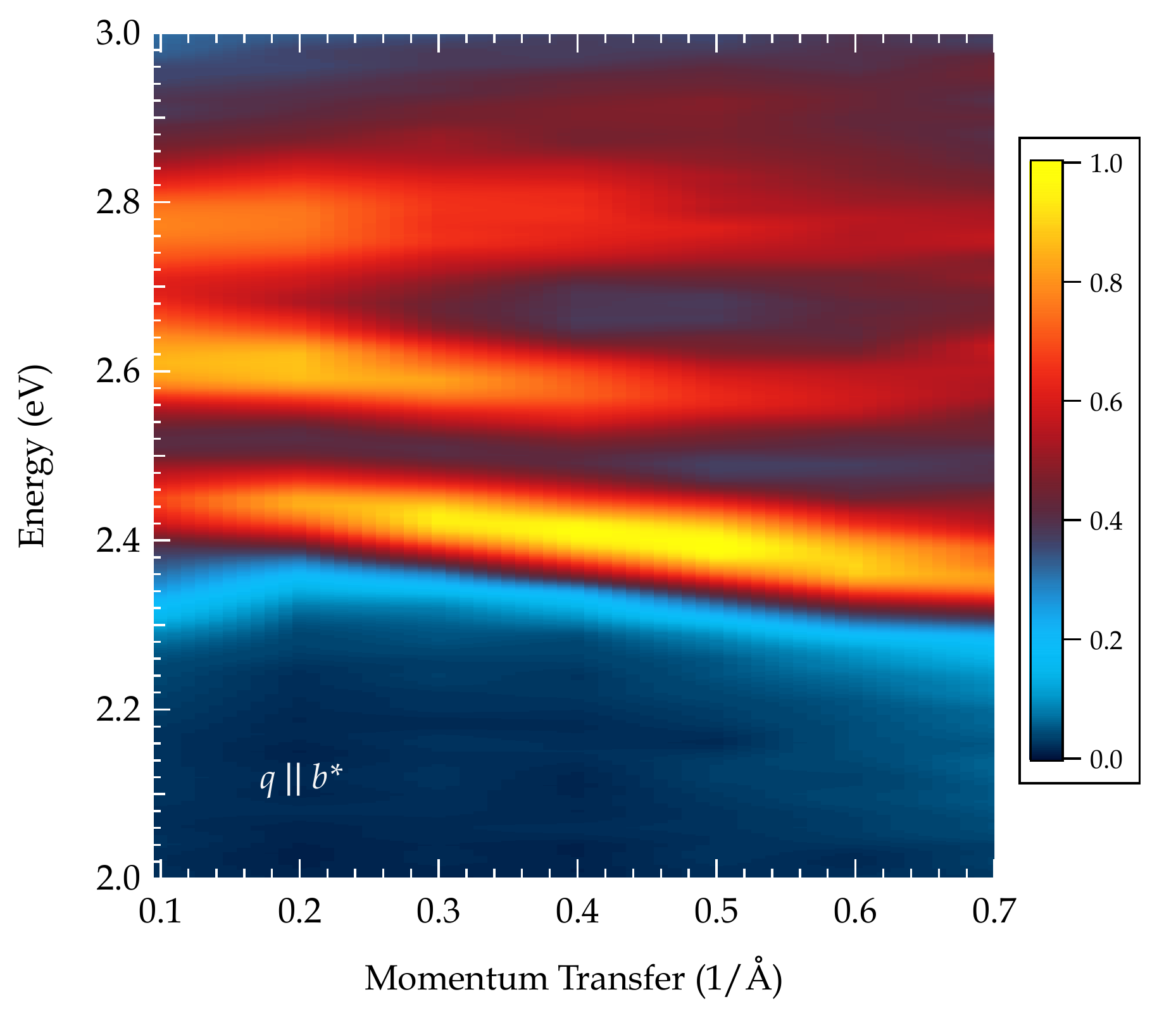}
\caption{Excitation spectra of tetracene as a function of momentum transfer parallel to the reciprocal lattice direction $b^*$. The data around 0.8\,\AA$^{-1}$ could not be determined accurately enough because of considerably enhanced multiple scattering in this region due to [010] Bragg reflection (cf. Fig.\,\ref{bragg}).}
\label{dispb}
\end{figure}

From previous momentum dependent studies of the electronic excitations in $\pi$ conjugated molecular crystals, it is well established that excitons in such materials can have a sizable dispersion giving rise to an anisotropic band structure \cite{Schuster2007,Knupfer2002,Roth2012}. Furthermore, the knowledge of this band structure can be essential in order to determine the microscopic character of the excitons. For pentacene, it has been demonstrated that the measured exciton dispersion of the lowest singlet excitons is not compatible with a pure Frenkel exciton character, but a substantial contribution of charge transfer excitons at lowest excitation energy must be considered \cite{Schuster2007,Roth2012,Yamagata2011,Cudazzo2012}. In Fig.\,\ref{dispa} and Fig.\,\ref{dispb} we depict the results of our momentum dependent studies for tetracene. In the upper panels we show selected excitation spectra and in the two lower panels the behavior is summarized in color plots. These data clearly demonstrate that also in tetracene, the lowest singlet exciton is characterized by a strong and anisotropic momentum dependence. For momentum transfers parallel to the $a^*$ direction (Fig.\,\ref{dispa}) we observe a clear upshift in energy with increasing momentum, while the spectral intensity decreases in this direction. This is qualitatively equivalent to the respective data for pentacene \cite{Roth2012}.

\par

\begin{figure}[h]
\includegraphics[width=.8\linewidth]{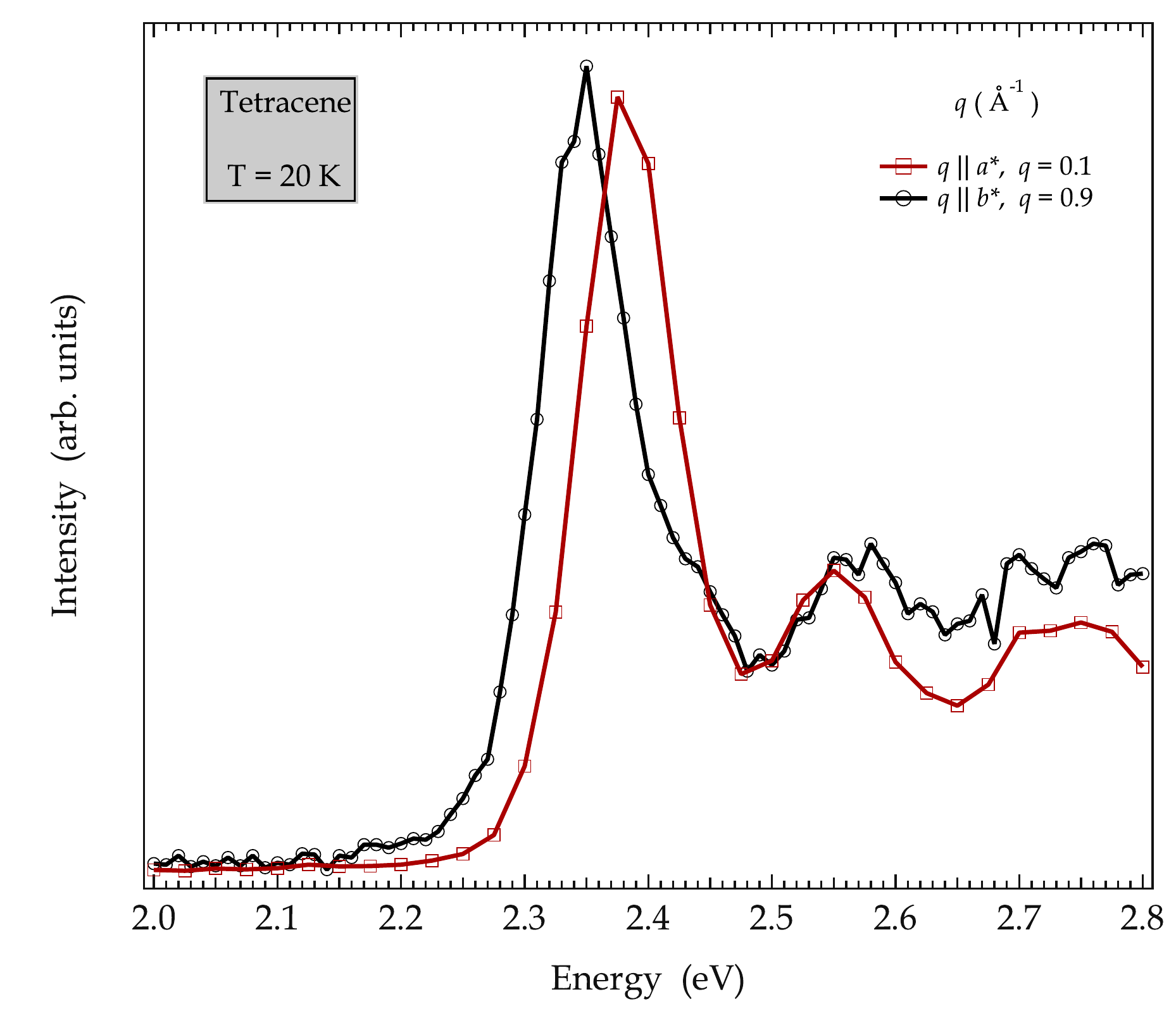}
\caption{Comparison of the excitation spectra for $q\|a^*$ in the optical limit ($q$ = 0.1\,\AA$^{-1}$) and for $q\|b^*$ at $q$ = 0.9\,\AA$^{-1}$.}
\label{vgls}
\end{figure}

For $q$ along the $b^*$ direction, the upper Davydov component shows a clear negative dispersion, while the shoulder at low energies vanishes. Also this behavior is reminiscent of that in pentacene.

\begin{figure}[t]
\includegraphics[width=0.9\linewidth]{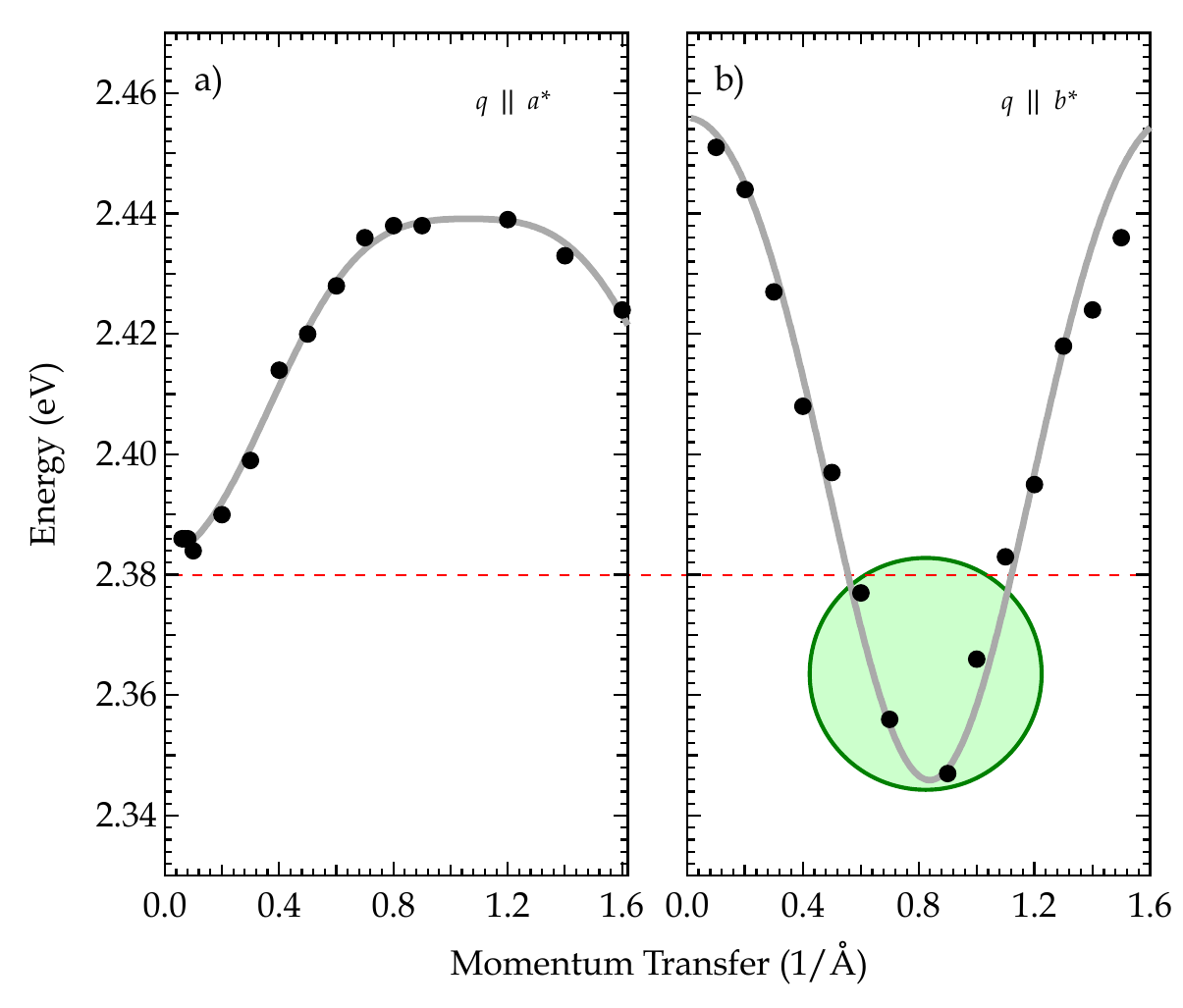}
\caption{Band structure of the singlet excitons in tetracene parallel to the two fundamental reciprocal lattice directions $a^*$ and $b^*$. The gray lines are intended as a guide to the eye. The green shaded area denotes the low energy exciton pocket observed for $q\|b^*$ in tetracene.}
\label{bandstr}
\end{figure}

There is, however, a surprising new observation - the exciton for $q$ along $b^*$ arrives at an energy lower than the lowest excitation energy for very small momentum values. At very small momentum ($q\sim$\,0\,\AA$^{-1}$), the data represent the so-called optical limit where they can directly be compared to the results of optical methods. This new observation is outlined in Fig.\,\ref{vgls}, where we directly compare the excitation spectra for $q\|a^*$ at a small momentum value of 0.1\,\AA$^{-1}$ to that for $q\|b^*$ and $q$ = 0.9\,\AA$^{-1}$. The analysis of the excitation maxima of the lowest excitons as a function of momentum complements this result and is shown in Fig.\,\ref{bandstr}. It becomes clear that the exciton band structure in tetracene is significantly anisotropic both in terms of the form of the dispersion curves and the total band width. For $q\|a^*$ the energy-momentum dependence E($q$) cannot be described by a simple cosine function as it was the case for pentacene \cite{Roth2012}, it considerably flattens out at the Brillouin zone boundary. Moreover, the total band width as seen in our data for tetracene of about 55\,meV is only half of the value measured for pentacene. In contrast, the curve for $q\|b^*$ is well represented by a simple cosine-like function and its band width is larger giving rise to the lowest singlet exciton energy at finite momentum. Note that the band width for $q\|b^*$ of about 105\,meV is very close to the corresponding value in pentacene.

The formation of a low energy exciton pocket as indicated in Fig.\,\ref{bandstr} can have important consequences for the photophysical behavior of tetracene. Excitations at higher energy might be able to relax into this pocket from where they cannot decay radiatively any more in a direct process due to momentum conservation. Exciton-phonon coupling, for instance, can support this relaxation. This would quench part of the fast photoluminescence and might be responsible for the differences between the behavior of tetracene and pentacene. Moreover, the exciton pocket could be associated with trap-like states or dark states as being discussed in the literature \cite{Tayebjee2013,Burdett2011,Burdett2010,Wilson2013}. In particular in view of singlet fission as a process to generate multiple excitons and to achieve high efficient organic photovoltaic cells a momentum dependence of the excitons as revealed in Fig.\,\ref{bandstr} might be a disadvantage, since the relaxation into the low energy pockets can compete to the fission process.

\par

\section{Summary}
To summarize, our results demonstrate that the lowest singlet excitons in tetracene are strongly momentum dependent giving rise to a particular, significantly anisotropic exciton band structure in the $a^*$, $b^*$ reciprocal lattice plane. Importantly, there is a low energy exciton pocket at finite momentum, which might explain a number of recent experimental results on the photophysical behavior of tetracene and the observed differences in comparison to its close relative pentacene. Our results emphasize the general importance of the exciton band structure for a thorough understanding of the photophysical behavior of organic semiconductors.

\acknowledgments
We thank M. Naumanm, R. H\"ubel and S. Leger for technical assistance. This work has been supported by the Deutsche Forschungsgemeinschaft (grant number KN393/14).

\end{document}